\documentclass[journal,transmag]{IEEEtran}

\usepackage{cite}
\usepackage{graphicx}
\usepackage{amssymb,amsmath}

\usepackage{tikz}

\begin{document}

\title{Magnetization reversal and direct observation of magnetic domains on FePt thin films}


\author{
	\IEEEauthorblockN{Augusto Roman\IEEEauthorrefmark{1,2}, Javier Gomez\IEEEauthorrefmark{3,4}, Alejandro Butera\IEEEauthorrefmark{3,4}, Paolo Vavassori\IEEEauthorrefmark{5,6} and Laura B. Steren\IEEEauthorrefmark{1,7}}
	\IEEEauthorblockA{\IEEEauthorrefmark{1}Instituto de Nanociencia y Nanotecnología, CNEA/CONICET, Nodo Constituyentes,\\ Av. Gral. Paz 1499, 1650, San Martín, Buenos Aires, Argentina.}
    \IEEEauthorblockA{\IEEEauthorrefmark{2}Departamento de Micro y Nanotecnología, CNEA, Av. Gral. Paz 1499, 1650, San Martín, Buenos Aires, Argentina}
	\IEEEauthorblockA{\IEEEauthorrefmark{3}Instituto de Nanociencia y Nanotecnología, CNEA/CONICET, Nodo Bariloche,\\ Av. Bustillo 9500, 8400, San Carlos de Barilo-che, Río Negro, Argentina}
    \IEEEauthorblockA{\IEEEauthorrefmark{4}Laboratorio de Resonancias Magnéticas, Centro Atómico Bariloche, CNEA, \\Av. Bustillo 9500, 8400, San Carlos de Bariloche, Río Negro, Argentina }
    \IEEEauthorblockA{\IEEEauthorrefmark{5}CIC nanoGUNE BRTA, 20018 Donostia-San Sebastián, Spain}
    \IEEEauthorblockA{\IEEEauthorrefmark{6}IKERBASQUE, Basque Foundation for Science, Plaza Euskadi 5, 48009 Bilbao, Spain}

    \IEEEauthorblockA{\IEEEauthorrefmark{7}Consejo Nacional de Investigaciones Científicas y Técnicas, Argentina}
}

\IEEEtitleabstractindextext{%
\begin{abstract}

In ferromagnetic thin films, the presence of an out-of-plane component of the magnetic anisotropy may induce a transition from planar to stripe-like magnetic domains above a critical thickness, $t_c$. Because of the changes in the domain structure, important changes in the magnetization reversal mechanisms are observed. We present the analysis of the magnetization reversal in FePt thin films, where this phenomenon is observed, through the experimental observation of magnetic domains by magneto-optic Kerr effect microscopy. The observed reversal mechanisms are strongly dependent on the thickness of the sample. For $t<t_c$, the reversal process is a combination of domain wall movement and rotation of domains. Meanwhile, for $t>t_c$, the stripe-domain structure determines the magnetization reversal.   

\end{abstract}

}

\maketitle

\pagestyle{empty}
\thispagestyle{empty}

\IEEEpeerreviewmaketitle

\section{Introduction}

\IEEEPARstart{T}{he} magnetization reversal mechanisms in thin films have been intensively studied in recent years due to their implications for understanding hysteresis loops and technological applications. The analysis of the reversal magnetization mechanisms and the possibility of controlling them are an essential input for technological applications \cite{harres2022magnetization,hazra2020magneto}. Moreover, controlling magnetic domains presents an exciting opportunity for advancing technology based on domain propagation, including logic circuits, racetrack memories, sensors, and RF devices \cite{garnier2020stripe}. Also, the knowledge of the magnetic components along the magnetization reversal process could be applied to the development of advance measurement techniques as spin-orbit torque measurements by optic methods \cite{celik2018vector, tsai2018spin}. Particularly, magnetic films with stripe-domain configuration offer diverse applications in new technologies, from optical to sensors and storage devices \cite{garnier2020stripe}.

FePt films magnetization may exhibit a stripe domain configuration with out-of-plane component depending on the film thickness. This system presents a transition between in-plane domains to stripe-like domains at a critical thickness $t_c\approx30$ nm \cite{leva2010magnetic,roman2023magnetization}. 

The mechanism of magnetization reversal in FePt films is still under discussion.  The reversal process above the critical thickness has been well described by micromagnetic simulations \cite{roman2023magnetization}. On the other hand, below the critical thickness, the reversal magnetization was described with a phenomenological model based on Neel phase theory \cite{roman2023magnetization,neel1960laws}. This model has been recently discussed because of its applicability to explain the magnetization reversal in different magnetic films \cite{hazra2020magneto, jamal2023interface, barwal2018growth, mohanta2019structural}. This model assumes that there are only two types of magnetic domains, e.g., two phases. The change in the proportion between these two phases defines the reversal process. 

In this work, we present experimental observations of the magnetization reversal process for films in both ranges in order to deepen the understanding of the reversal process of this system. 

\section{Experimental details}
\begin{figure}[ht]
\includegraphics[width=\columnwidth]{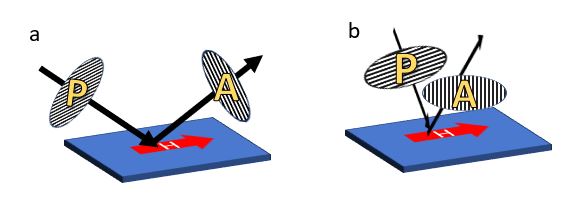}
\caption{Magneto-Optical Kerr Effect (MOKE) Experiment Scheme featuring Polarizer (P), Analyzer (A), and Applied Magnetic Field (H). (a) Longitudinal MOKE; (b) Transverse MOKE.\label{Experimental_MOKE}	 }
\end{figure}

We studied the magnetism of a series of FePt thin films fabricated at room temperature by DC magnetron sputtering on single crystal SrTiO$_3$ substrates. The chamber was pumped down to a base pressure of $1\times10^{-6}$ Torr, and the films were sputtered at an Ar pressure of 2.6 mTorr. A power of 20 W and a target-substrate distance of about 10 cm were used. The sputtering rate for the FePt deposition was 0.19 $\frac{\mathrm{nm}}{\mathrm{s}}$. A 4 nm-thick Ru layer capped the samples to prevent oxidation. The films were grown with the following thicknesses: 10 nm, 20 nm, 40 nm, and 60 nm. FePt films primarily grow in the disordered phase and exhibit a polycrystalline structure. As was previously reported on silicon substrates, the out-of-plane magnetic anisotropy is related to strains induced during the fabrication process \cite{leva2010magnetic, roman2023magnetization}. A more detailed structural characterization of the FePt films deposited on STO substrates will be discussed elsewhere \cite{romanmagneto}.

\begin{figure}[ht]
\begin{tikzpicture}
\draw (1.5, 0) node[inner sep=0] 
{
\includegraphics[width=0.7\columnwidth]{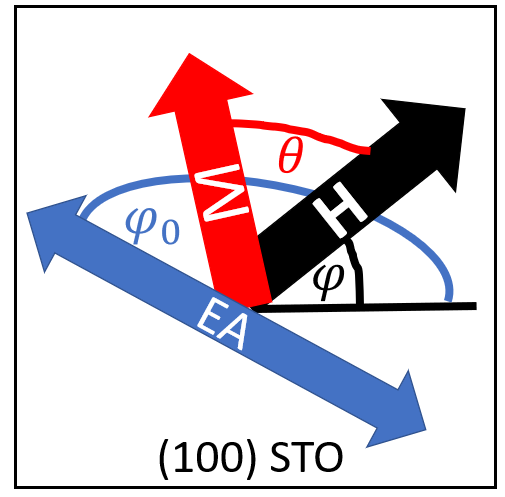}
};
\draw[font=\Large] (-2.6, 3) node{(a)};
\end{tikzpicture} 
\begin{tikzpicture}
\draw (0, 0) node[inner sep=0] {\includegraphics[width=\columnwidth]{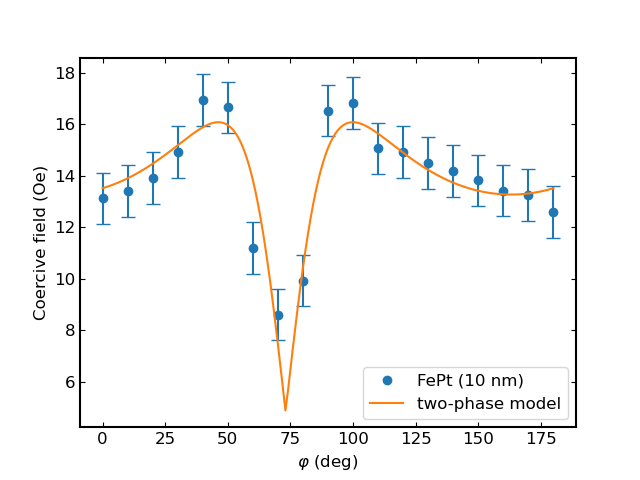}};
\draw[font=\Large] (-4, 2.7) node {(b)};
\end{tikzpicture} 

\caption{(a) Scheme of angles regarding magnetization (M), applied magnetic field (H), FePt films easy axis (EA) and (100) STO crystalline direction. (b) Coercive field angular dependence from a 10 nm FePt film \label{FePt10nm_Hc_angular}	 }
\end{figure}

The magnetism of the films was analyzed by applying the magnetic field in the plane of the film. The angle $\varphi$ is defined as the angle between the edge of the substrate -(100) STO crystal axis- and the direction of the magnetic field ($H$). Magneto Optical Kerr Effect (MOKE) hysteresis loops and magnetic domain images were measured using a Kerr effect microscope with polarization analysis control, operated in both longitudinal and transversal configurations (Fig. \ref{Experimental_MOKE}). These configurations allowed us to obtain the parallel and transversal components of the magnetization, respectively. The out of plane component of the magnetization at remanence was studied by Magnetic Force Microscopy (MFM).

\section{Results and discussion}

The analysis of the measurements varying $\varphi$ let us identify an in-plane uniaxial anisotropy for the thinner films (10 nm and 20 nm). The maximum in the remanence angular dependence indicates that the easy axis is at $\varphi=160^0$ and the hard axis is at $\varphi=70^0$. An isotropic response to magnetic field was observed for films thicker than the critical thickness (40 nm and 60 nm). 

From MFM measurements, we observed that the thinner films didn't present any magnetic contrast, while thicker films (40 nm and 60 nm) present stripe domains. These results indicate that the critical thickness is between 20 nm and 40 nm, in agreement with our previous observations on FePt films deposited on silicon \cite{roman2023magnetization, leva2010magnetic}. 

\subsection{Magnetization reversal for $t<t_c$}

\begin{figure}[hb]
\begin{tikzpicture} 
\draw (0, 0) node[inner sep=0] {\includegraphics[width=\columnwidth]{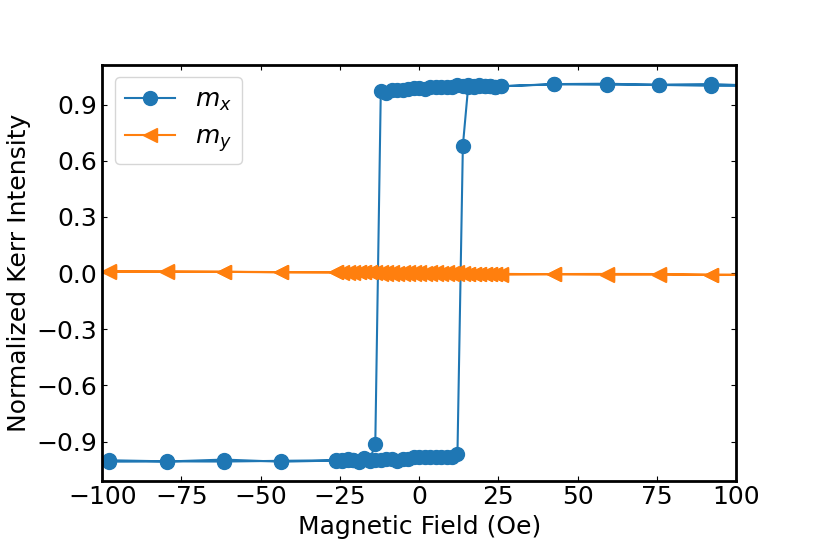}};
\draw[font=\Large] (3, -1.75) node {(a)};
\end{tikzpicture} 
\begin{tikzpicture} 
\draw (0, 0) node[inner sep=5] {    \includegraphics[width=0.9\columnwidth]{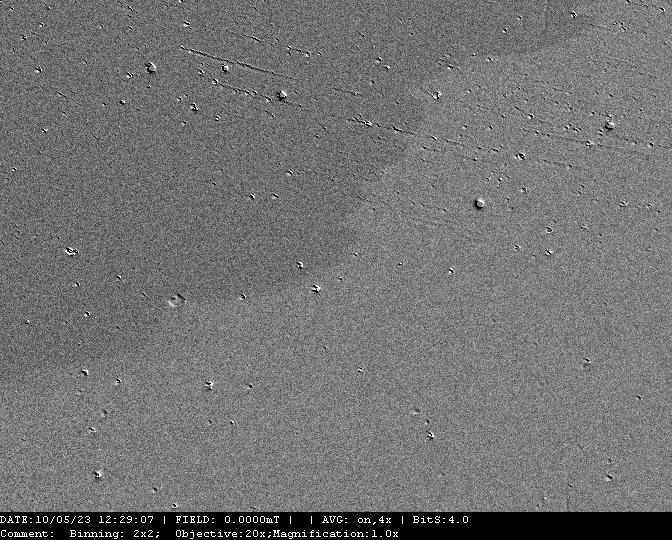}};
    \draw[white, font=\Large] (-3,-2) node{100$\mu$m};
    \draw[font=\Large, white] (3, -2.0) node {(b)};
    \draw[white, line width=1mm] (-3.75,-2.25)--(-1.94,-2.25);
    \draw[white, line width=1mm,->] (2,-0.2)--(-2,-0.2);
    \draw[black, line width=1mm,<->] (-3,2.1)--(-2,2.1);
    \draw[black, font=\Large](-2.5,2.5) node{\textbf{$m_x$}};
    \draw[white, font=\Large] (0,0.25) node{\textbf{H}};
\end{tikzpicture}
\caption{ MOKE measurements of a 10 nm FePt film with the magnetic field applied parallel at $\varphi = 160^0$
(easy axis): (a) Normalized magnetization curves
parallel ($m_x$) and perpendicular ($m_y$) to the applied magnetic field (H). (b) Domains image captured at the coercive field ($H_c=13 Oe$) .\label{FePt10nm_EA}	}
\end{figure}

The angular dependence of the coercive field is a footprint of the magnetization reversal mechanism \cite{oh2005crystallographic}.  The two-phase model described by Eq. \ref{twophasemodel}\cite{suponev1996angular} is the best fit for our experimental data (Fig. \ref{FePt10nm_Hc_angular}).

\begin{multline}
    H_C\left(\varphi\right) =max\left[ \frac{H_0\cos\left(\varphi-\varphi_0 \right)}{\frac{1}{y}\sin^{2}\left(\varphi-\varphi_0\right)+\cos^{2}\left(\varphi-\varphi_0\right)},H_{pol}\right],
    \\
    y =\frac{N_{A}+N_{x}}{N_{y}}
    \label{twophasemodel}
\end{multline}
$N_x$, $N_y$, and $N_z$ are the demagnetizing factors of the ellipsoid along its main axes, being $N_x=N_z$. $N_A$ is an effective demagnetizing factor that takes into account the contributions of anisotropies other than shape anisotropy. The coercive field is not zero at the hard axis $H_c(\varphi=70^o)$, this minimum in the coercive field ($H_{pol}$) is attributed to the grain easy axis distribution due to the polycrystalline nature of the sample. $\varphi$ is the angle between (100)-STO crystalline direction and the direction of the applied magnetic field and $\varphi_0=160^0$ is the angle between (100)-STO crystal axis and the easy axis of the FePt film.

In Fig. \ref{FePt10nm_EA} (a), magnetization loops measured with the magnetic field applied along the easy-axis are shown. The magnetization curve measured in the longitudinal configuration ($m_x$) is a square loop with an abrupt transition at the coercive field, while there is no magnetic signal in the transversal configuration ($m_y$).

\begin{figure}[ht]
\begin{tikzpicture} 
\draw (0, 0) node[inner sep=0] {\includegraphics[width=\columnwidth]{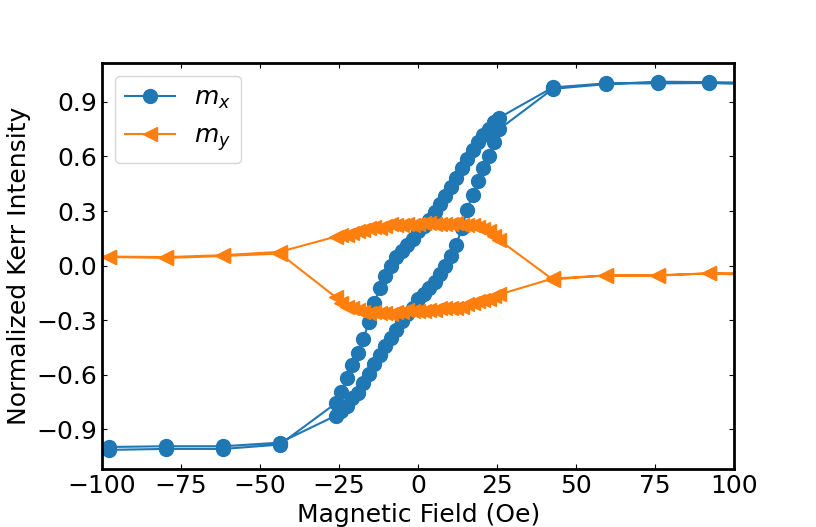}};
\draw[font=\Large] (3, -1.75) node {(a)};
\end{tikzpicture} 
\begin{tikzpicture} 
\draw (0, 0) node[inner sep=5] {    \includegraphics[width=0.9\columnwidth]{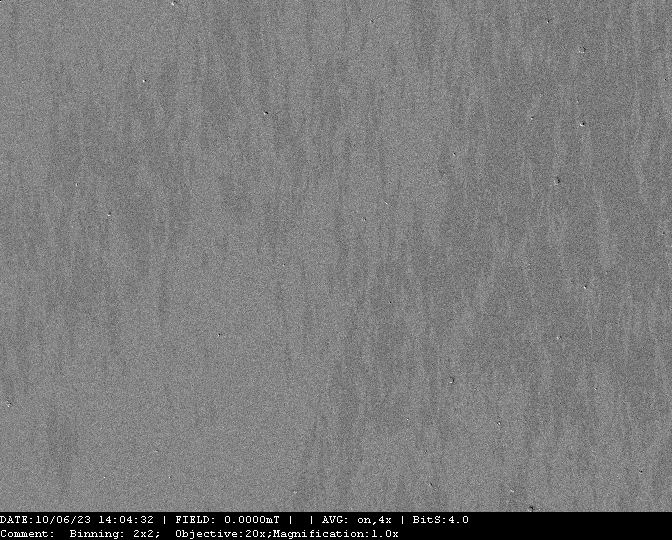}};
    \draw[white, font=\Large] (-3,-2) node{100$\mu$m};
    \draw[font=\Large, white] (3, -2.0) node {(b)};
    \draw[white, line width=1mm] (-3.75,-2.25)--(-1.94,-2.25);
    \draw[white, line width=1mm,->] (2,-0.2)--(-2,-0.2);
    \draw[white, font=\Large] (0,0.25) node{\textbf{H}};
    \draw[black, font=\Large](-2.5,2) node{\textbf{$m_y$}};
    \draw[lightgray, line width=2.5mm,-latex] (-3,2)--(-3,3.2);
    \draw[black, line width=2.5mm,latex-] (-3,0.7)--(-3 ,2);
\end{tikzpicture}
\caption{ MOKE measurements of a 10 nm FePt film with the magnetic field applied parallel at $\varphi = 70^0$ (hard axis): (a) Normalized magnetization curves parallel ($m_x$) and perpendicular ($m_y$) to the applied magnetic field (H). (b) Domains image captured at the coercive field ($H_c=8\ \mathrm{Oe}$). \label{FePt10nm_HA}	}
\end{figure}

The image obtained by MOKE microscopy near to the coercive field is shown in Fig. \ref{FePt10nm_EA} (b). The image shows two big domains with opposite directions of magnetization. The size of one of those domains increases through domain wall movement throughout the magnetization reversal process until the magnetization is completely reversed. In this case, the coercive field, understood as the magnetic field where the magnetization reverses, agrees with the switching field where the nucleation of domains and a sharp change in the magnetization take place.

\begin{figure}[ht]
\begin{tikzpicture} 
\draw (0, 0) node[inner sep=0] {\includegraphics[width=\columnwidth]{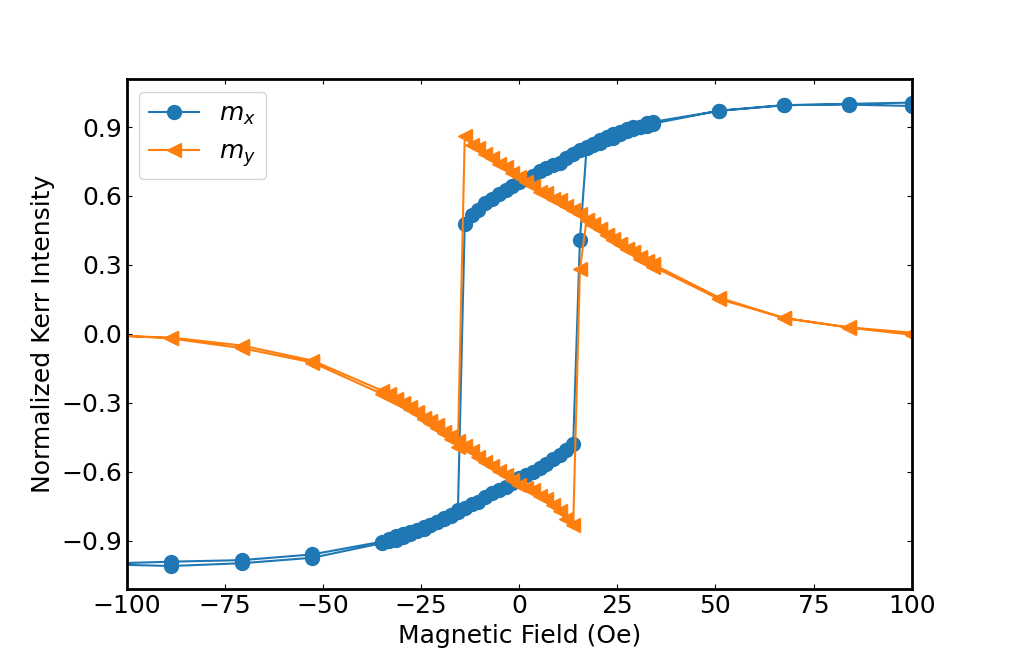}};
\draw[font=\Large] (3, -1.75) node {(a)};
\end{tikzpicture} 
\begin{tikzpicture} 
\draw (0, 0) node[inner sep=0] {    \includegraphics[width=\columnwidth]{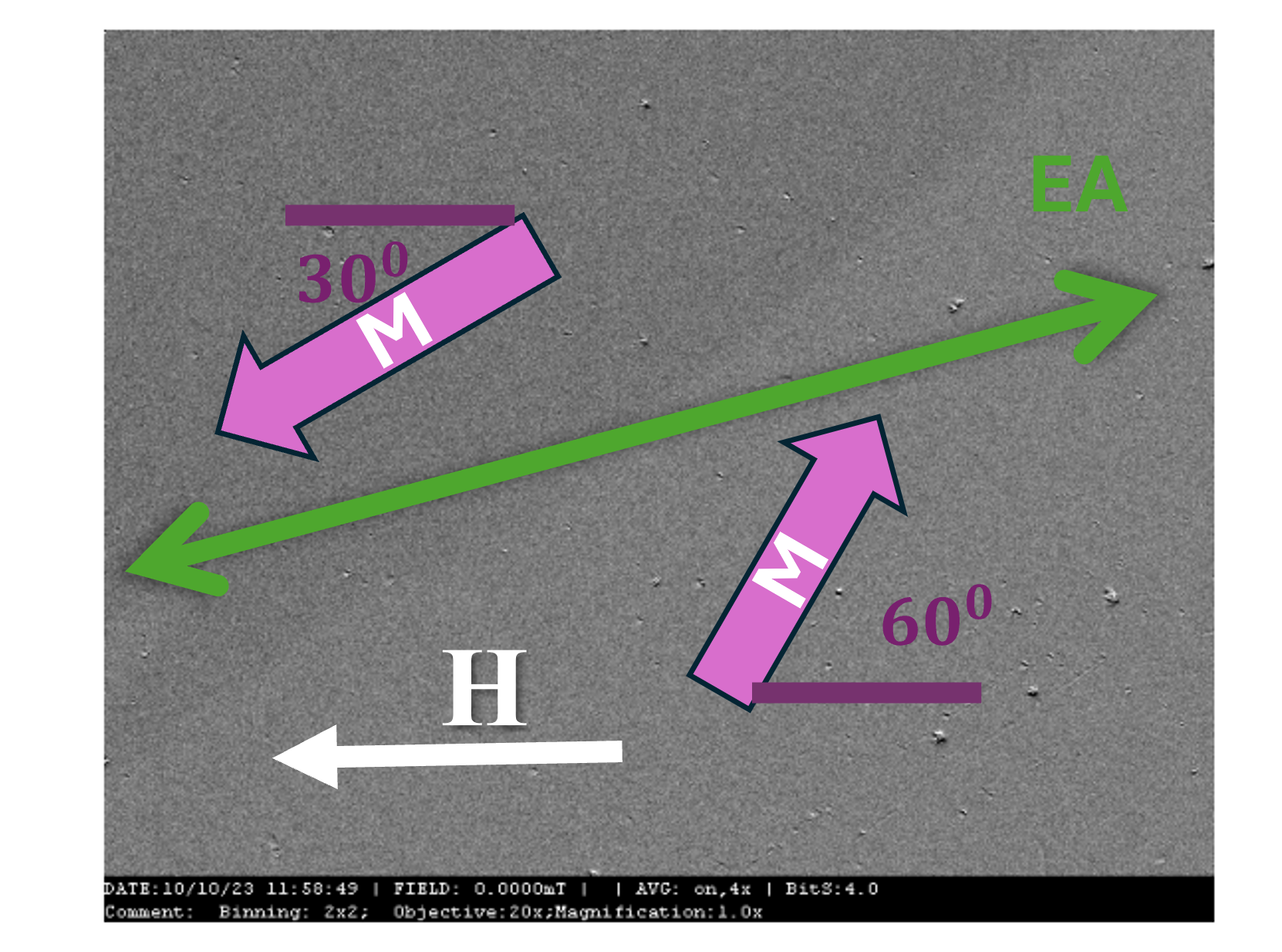}};
\draw[font=\Large, white] (3, -2.2) node {(b)};
\end{tikzpicture}
\caption{ MOKE measurements from a 10 nm FePt film with the magnetic field applied at $\varphi = 30^\circ$: (a) Normalized magnetization curves parallel ($m_x$) and perpendicular ($m_y$) to the applied magnetic field (H). (b) Scheme of magnetization (M) angles with the applied magnetic field (H) measured at the coercive field ($H=H_c=15 \mathrm{Oe}$). The arrows indicate the hard (HA) and easy (EA) axes.\label{FePt10nm_30deg}}
\end{figure}

In Fig. \ref{FePt10nm_HA} (a) we present the magnetization loops measured with the applied field parallel to the hard-axis. For magnetic fields below the saturation field ($H<H_s$), the transversal component increases as the magnetic field decreases, reaching a maximum at the coercive field, and the squared sum of both magnetizations is smaller than one. This suggests the formation of domains. The images obtained by MOKE microscopy (Fig. \ref{FePt10nm_HA} (b)) confirm the formation of small domains in the transversal direction. 

We analyzed the magnetization profiles extracted from MOKE and we observed a progressive change in the magnetization components ($m_x$ and $m_y$) with the magnetic field. Specifically, $m_x$ reverses gradually and without magnetic contrast spatial dependence during the formation of opposite magnetic domains in the transverse component ($m_y$) of the magnetization. These facts suggest that magnetization reversal, when the magnetic field is applied along the hard axis, occurs through the rotation of the domains.

The local maxima in the angular dependence of the coercive field (Fig. \ref{FePt10nm_Hc_angular}) defines a region between 40$^0$ and 100$^0$ that contains the hard axis. In the following, we are going to compare the magnetization reversal when the magnetic field is applied at an angle outside this region or inside it.

In Figure \ref{FePt10nm_30deg}, we present the normalized magnetization curves from the 10 nm FePt films with the magnetic field applied at $\varphi= 30^\circ$. The relationship between $m_x$ and $m_y$ in the magnetization curves suggests a rotation of the magnetization. The MOKE images didn't show magnetic contrast except in images captured at the coercive field ($H_c$), which confirm that for magnetic fields different from $H_c$, the film is a single magnetic domain. Based on this observation, we calculated $\theta$, the angle between the magnetic field and the magnetization. At the remanence the magnetization is aligned with the easy axis which is at $\theta=50^0$. Between remanence and the coercive field the angle increases, preserving the single-domain state reaching a maximum of $\theta_{max}=60^0$ at the coercive field, where the magnetization reverses trough domain wall movement. The angle between applied magnetic field and the magnetization after the domain wall displacement is $30^0$.  In Fig. \ref{FePt10nm_30deg} (b) we schematize the magnetic domains at the coercive field.

\begin{figure}[ht]
\includegraphics[width=\columnwidth]{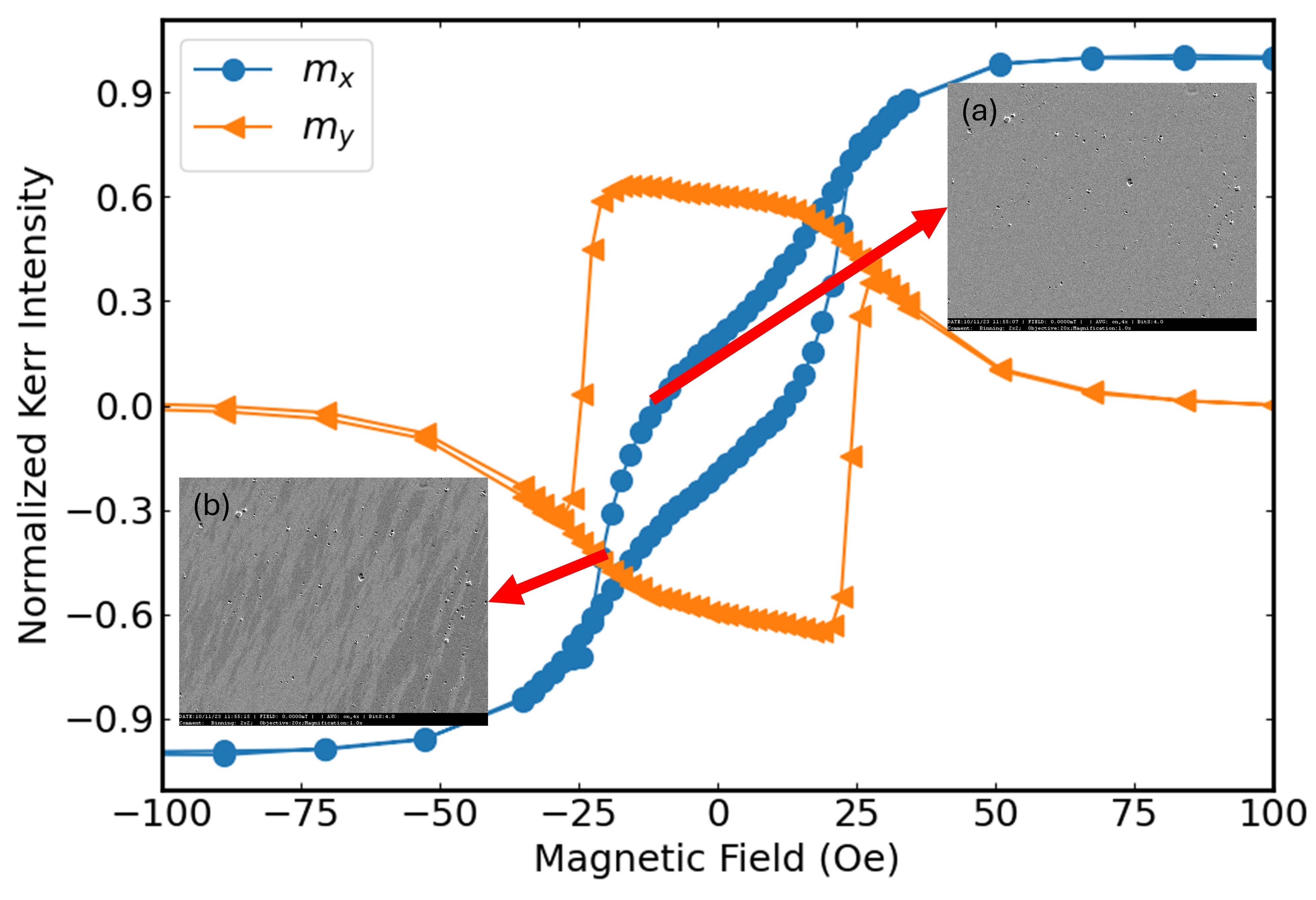}
\caption{ MOKE measurements from a 10 nm FePt film with the magnetic field applied at $\varphi = 60^\circ$: INSET: (a) Domain images captured at the coercive field ($H_c= 11 \ \mathrm{Oe}$). (b) Domain images captured at the switching field ($H_{sw}=24\ \mathrm{Oe} >H_c$)\label{FePt10nm_60deg}}
\end{figure}

For $\varphi=60^0$, the magnetization curves suggest that magnetization rotates coherently until a switching field at $H_{sw}=24 \ \mathrm{Oe}$. As we present in the Fig. \ref{FePt10nm_60deg} (a), at the coercive field, the MOKE image doesn't show spatially dependent magnetic contrast, which indicates that the magnetization reverses by coherent rotation. Domain nucleation is at $H_{sw}$, as we show in Fig. \ref{FePt10nm_60deg} (b).

Considering these results, we can see that the maximum in the coercive field defines two regions. One that includes the easy axis, where the magnetization reversal is mainly by domain wall movement and a second one that includes the hard axis, where the reversal magnetization takes place mainly by rotation of domains. 

Our results from films thiner than FePt critical thickness are in agreement with the two-phase model, which has been associated with the combination of rotation of domains and domain wall movement in the magnetization reversal\cite{suponev1996angular}. These results present experimental evidence of the role of each reversal mechanism in the experimental system and its relationship with the angular dependence of coercivity.

\subsection{Magnetization reversal for $t>t_c$}
\begin{figure}[ht]
\begin{tikzpicture} 
\draw (0, 0) node[inner sep=0] {\includegraphics[width=\columnwidth]{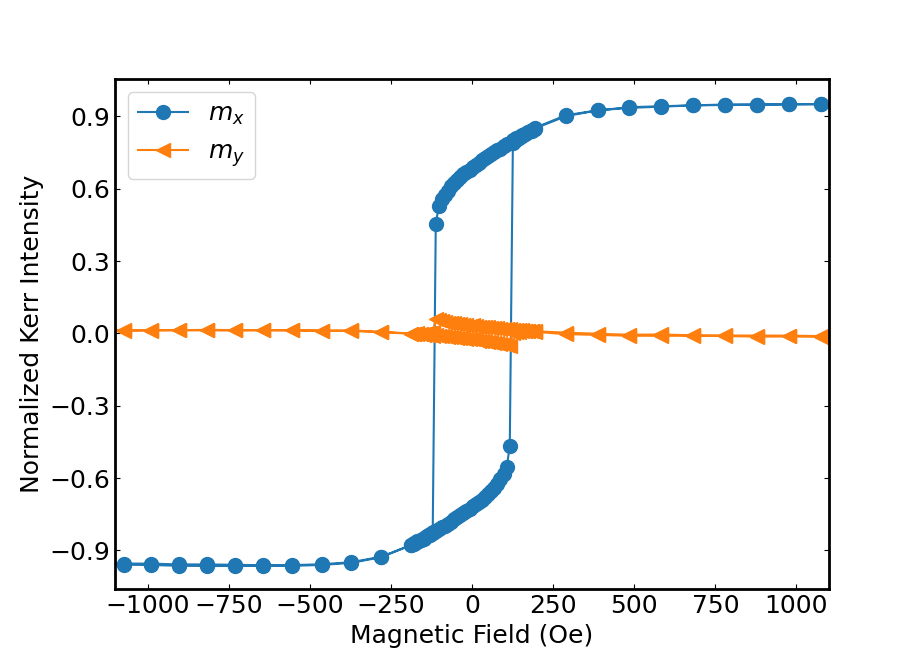}};
\draw[font=\Large] (3, -1.75) node {(a)};
\end{tikzpicture} 
\begin{tikzpicture} 
\draw (0, 0) node[inner sep=5] {    \includegraphics[width=0.505\columnwidth]{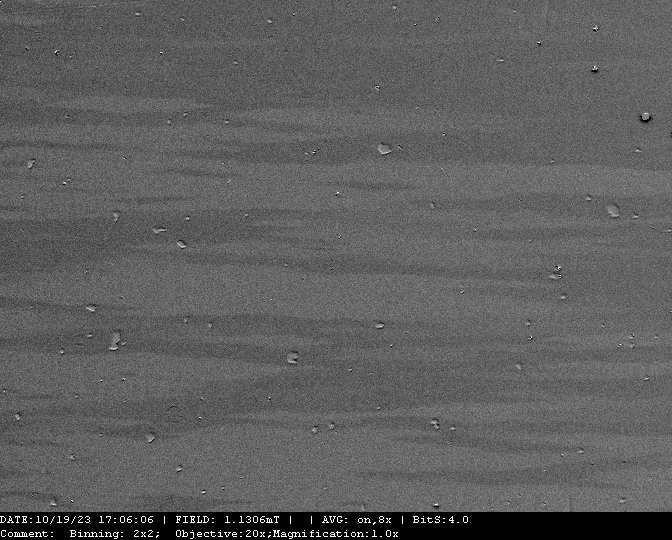} 

\includegraphics[width=0.41\columnwidth]{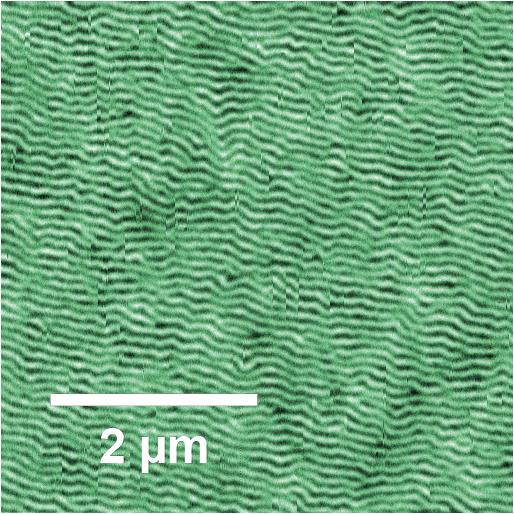}};
    \draw[white, font=\large] (-3.2,-1.3) node{100$\mu$m};
    \draw[font=\Large, white] (-0, -1.3) node {(b)};
    \draw[font=\Large, white] (3.6, -1.3) node {(c)};
    \draw[white, line width=1mm] (-3.75,-1.0)--(-2.84,-1.0);
    \draw[white, line width=1mm,->] (-1,-0.2)--(-3,-0.2);
    \draw[white, line width=1mm,->] (3.25,-0.2)--(1.25,-0.2);
    \draw[white, font=\Large] (-1.75,0.25) node{\textbf{H}};
    \draw[white, font=\Large] (2.5,0.25) node{\textbf{H}};
    \draw[gray, line width=2.5mm,-latex] (-2.5,1.1)--(-1.3,1.1);
    \draw[black, line width=2.5mm,latex-] (-3.7,1.1)--(-2.5 ,1.1);
    \draw[black, font=\large](-2.5,1.5) node{\textbf{$m_x$}};
\end{tikzpicture}
\caption{ MOKE measurements of a 40 nm FePt film with magnetic field applied parallel to the plane of the film: (a) Normalized magnetization curves
parallel ($m_x$) and perpendicular ($m_y$) to the applied magnetic field (H). (b)
Domains image captured at the coercive field. (c) MFM image measured at remanence after aplying a 2000 Oe  magnetic field (H). \label{FePt40nm}}
\end{figure}

In Fig. \ref{FePt40nm} (a), we present the magnetization curve from a 40 nm FePt film. The coercive field is larger in comparison to that of thinner films, and the magnetization presents a linear dependence with the magnetic field close remanence. These characteristics agree with the typical magnetization curves obtained for stripe-domain thin films. A small value of the transverse component of the magnetization ($m_y$) agrees with the hypothesis of the coherent rotation of magnetic moments inside each magnetic stripe \cite{roman2023magnetization}.  

The in-plane magnetic domains from the 40 nm FePt film measured at the coercive field are shown in Fig. \ref{FePt40nm} (b). Striped out of plane domains where observed at remanence Fig. \ref{FePt40nm} (c). The size of stripe-domains is 45 nm while in-plane domains are tens of micrometers. The formation of in-plane domains at the coercive field suggests that besides the rotation of magnetic moments of each stripe is coherent, the inversion of the stripes is not simultaneous. 

\section{Conclusions}
The magnetization reversal was directly observed through MOKE microscopy for FePt thin films of different thicknesses. The reversal mechanisms are strongly dependent on the thickness of the sample. Below the critical thickness, the reversal process is a combination of domain wall movement and rotation of domains. The dominant reversal mechanism depends on the direction of the applied field. The local maxima on the coercive field angular dependence defines two angular ranges with different reversal mechanism. When the magnetic field is applied at angles close to the hard axis the main reversal mechanism is the rotation of the domains, while when the field is applied at angles close to the easy axis, the magnetization reverses through domain wall movement. Above the critical thickness, the experimental results agree with the hypothesis of coherent rotation of magnetic moments inside each stripe that was previously deduced from micromagnetic simulations.

\section{Acknowledgements}
We acknowledge the financial support of the European Commission by the H2020-MSCA RISE project ULTIMATE-I (Grant No 101007825).

\bibliographystyle{IEEEtran}
\bibliography{IEEEabrv,refs}

\begin{thebibliography}{10}
\providecommand{\url}[1]{#1}
\csname url@samestyle\endcsname
\providecommand{\newblock}{\relax}
\providecommand{\bibinfo}[2]{#2}
\providecommand{\BIBentrySTDinterwordspacing}{\spaceskip=0pt\relax}
\providecommand{\BIBentryALTinterwordstretchfactor}{4}
\providecommand{\BIBentryALTinterwordspacing}{\spaceskip=\fontdimen2\font plus
\BIBentryALTinterwordstretchfactor\fontdimen3\font minus \fontdimen4\font\relax}
\providecommand{\BIBforeignlanguage}[2]{{%
\expandafter\ifx\csname l@#1\endcsname\relax
\typeout{** WARNING: IEEEtran.bst: No hyphenation pattern has been}%
\typeout{** loaded for the language `#1'. Using the pattern for}%
\typeout{** the default language instead.}%
\else
\language=\csname l@#1\endcsname
\fi
#2}}
\providecommand{\BIBdecl}{\relax}
\BIBdecl

\bibitem{harres2022magnetization}
A.~Harres, T.~Mallmann, M.~Gamino, M.~A. Correa, A.~D. Viegas, and R.~B. da~Silva, ``Magnetization reversal processes in amorphous cofeb thin films,'' \emph{Journal of Magnetism and Magnetic Materials}, vol. 552, p. 169135, 2022.

\bibitem{hazra2020magneto}
B.~K. Hazra, S.~Kaul, S.~Srinath, Z.~Hussain, V.~R. Reddy, and M.~M. Raja, ``Magneto-optical kerr microscopy investigation of magnetization reversal in co2fesi heusler alloy thin films,'' \emph{AIP Advances}, vol.~10, no.~6, 2020.

\bibitem{garnier2020stripe}
L.-C. Garnier, M.~Marangolo, M.~Eddrief, D.~Bisero, S.~Fin, F.~Casoli, M.~G. Pini, A.~Rettori, and S.~Tacchi, ``Stripe domains reorientation in ferromagnetic films with perpendicular magnetic anisotropy,'' \emph{Journal of Physics: Materials}, vol.~3, no.~2, p. 024001, 2020.

\bibitem{celik2018vector}
H.~Celik, H.~Kannan, T.~Wang, A.~R. Mellnik, X.~Fan, X.~Zhou, R.~Barri, D.~C. Ralph, M.~F. Doty, V.~O. Lorenz \emph{et~al.}, ``Vector-resolved magnetooptic kerr effect measurements of spin--orbit torque,'' \emph{IEEE Transactions on Magnetics}, vol.~55, no.~1, pp. 1--5, 2018.

\bibitem{tsai2018spin}
T.-Y. Tsai, T.-Y. Chen, C.-T. Wu, H.-I. Chan, and C.-F. Pai, ``Spin-orbit torque magnetometry by wide-field magneto-optical kerr effect,'' \emph{Scientific reports}, vol.~8, no.~1, p. 5613, 2018.

\bibitem{leva2010magnetic}
E.~S. Leva, R.~Valente, F.~M. Tabares, M.~V. Mansilla, S.~Roshdestwensky, and A.~Butera, ``Magnetic domain crossover in fept thin films,'' \emph{Physical Review B}, vol.~82, no.~14, p. 144410, 2010.

\bibitem{roman2023magnetization}
A.~Rom{\'a}n, A.~G. Lopez~Pedroso, K.~Bouzehouane, J.~Gomez, A.~Butera, M.~Aguirre, M.~M. Soares, C.~Garcia, and L.~B. Steren, ``Magnetization reversal modes and coercive field dependence on perpendicular magnetic anisotropy in fept thin films,'' \emph{Journal of Physics D: Applied Physics}, 2023.

\bibitem{neel1960laws}
L.~N{\'e}el, R.~Pauthenet, G.~Rimet, and V.~Giron, ``On the laws of magnetization of ferromagnetic single crystals and polycrystals. application to uniaxial compounds,'' \emph{Journal of Applied Physics}, vol.~31, no.~5, pp. S27--S29, 1960.

\bibitem{jamal2023interface}
M.~S. Jamal, P.~Gupta, I.~Sergeev, O.~Leupold, and D.~Kumar, ``Interface-resolved study of magnetism in mgo/fecob/mgo trilayers using x-ray standing wave techniques,'' \emph{Physical Review B}, vol. 107, no.~7, p. 075416, 2023.

\bibitem{barwal2018growth}
V.~Barwal, S.~Husain, N.~Behera, E.~Goyat, and S.~Chaudhary, ``Growth dependent magnetization reversal in co2mnal full heusler alloy thin films,'' \emph{Journal of Applied Physics}, vol. 123, no.~5, 2018.

\bibitem{mohanta2019structural}
M.~Mohanta, S.~Parida, A.~Sahoo, Z.~Hussain, M.~Gupta, V.~Reddy, and V.~Medicherla, ``Structural and magnetic properties of coni surface alloys,'' \emph{Physica B: Condensed Matter}, vol. 572, pp. 105--108, 2019.

\bibitem{romanmagneto}
A.~Román, A.~L. Pedroso, L.~Steren, J.~Gomez, A.~Butera, I.~Neckel, and M.~M. Soares, ``Thermal effects on the magneto-elastic coupling across fept/batio$_3$ interfaces,'' \emph{In elaboration}.

\bibitem{oh2005crystallographic}
D.~Oh and J.~K. Park, ``Crystallographic texture and angular dependence of coercivity of ordered copt thin film,'' \emph{Journal of Applied Physics}, vol.~97, no.~10, p. 10N105, 2005.

\bibitem{suponev1996angular}
N.~Suponev, R.~Grechishkin, M.~Lyakhova, and Y.~E. Pushkar, ``Angular dependence of coercive field in (sm, zr)(co, cu, fe) z alloys,'' \emph{Journal of Magnetism and Magnetic Materials}, vol. 157, pp. 376--377, 1996.

\end{thebibliography}

\end{document}